\documentclass[11pt]{article}
\usepackage{a4}
\usepackage{cite}
\usepackage{graphicx}
\usepackage{amssymb}
\usepackage{algorithmicx}
\usepackage{algpseudocode}
\usepackage[ruled]{algorithm}
\renewcommand{\algorithmiccomment}[1]{\hspace*{4mm}{\em //~#1}}

\newtheorem{definition}{Definition}
\newtheorem{theorem}{Theorem}
\newtheorem{lemma}{Lemma}
\newtheorem{claim}{Claim}

\def\QED{\ifvmode\vskip-\lastskip\nobreak\leavevmode\fi\hbox{}\hskip1cm plus 1fill\hbox{}\nobreak\hskip0cm plus1fill\nobreak$\Box$}
\newenvironment{proof}{{\em Proof:}~ }{\QED \vskip 2mm}
\newenvironment{proofsk}{{\em Sketch of the proof:}~ }{\QED \vskip 2mm}

\newcommand{\bm}[1]{\mbox{\boldmath{$#1$}}}

\newcounter{pako}

\def\setnumber#1#2{\expandafter\ifx\csname r@#2\endcsname\relax\else\setcounter{#1}{-1}\addtocounter{#1}{\ref{#2}}\fi}

\begin{document}
\title{Rapid Almost-Complete Broadcasting in Faulty Networks\thanks{The research has been
supported by grant APVV-0433-06.}}

\author{Rastislav Kr\'alovi\v{c}\footnotemark[3] \and Richard Kr\'alovi\v{c}\footnotemark[3] \footnotemark[4]\and \\[2mm]
\footnotemark[3] Department of Computer Science, Comenius University,\\
Mlynsk\'a dolina, 84248 Bratislava, Slovakia.
\\[2mm]
\footnotemark[4] Department of Computer Science, ETH Zurich, Switzerland.
}

\maketitle

\begin{abstract}
This paper studies the problem of broadcasting in synchronous point-to-point networks, where one
initiator owns a piece of information that has to be transmitted to all other vertices as fast as possible. 
The model of fractional dynamic
faults with threshold is considered: in every step either a fixed number $T$, or 
a fraction $\alpha$, of sent messages can be lost depending on which quantity is larger. 

As the main result we show that in complete graphs and hypercubes it is possible to inform all
but a constant number of vertices, exhibiting only a logarithmic slowdown, i.e. in time $O(D\log n)$
where $D$ is the diameter of the network and $n$ is the number of vertices.

Moreover, for complete graphs under some additional conditions 
(sense of direction, or $\alpha<0.55$)
the remaining constant number of vertices can be informed in the same time, i.e. $O(\log n)$.
\end{abstract}

\section{Introduction}

Fault tolerance has been a crucial issue in the distributed computing since its
beginnings \cite{B73,D74,BH86,ChM84,LPS80,FLP85}. 
Because a typical distributed system is designed to contain a large number of individual components,
attention must be paid to the fact that, even if the failure probability of a single component is negligible,
the probability that some components fail may be high.
There are numerous ways how to cope with failures, using either
probabilistic or deterministic approaches. In the probabilistic setting, it is supposed that
a failure probability of each component follows some probability
distribution \cite{BDP97,CDP91,DP92,P91,PP05}. 
Failures of individual
components are usually assumed to be independent random events. The goal is to design
algorithms and protocols that perform well with high probability if the failures follow the 
conjectured distribution. 

The deterministic approach, which is pursued also in this paper, copes with failures in a different way. 
Instead of considering a failure probability distribution for each individual component, 
algorithms and protocols are designed to perform well in the worst case, under some a-priori constraints on the 
failure behavior.
\cite{CDP94,DV99,DV04,LN00,F92,BH94,AGHK96,GP98,KKR03,DKKS06,SW90}.
These constraints may take the form of considering only computations with
a limited overall number of faults \cite{F92,AGHK96},
limited number of faults during any single computation step
\cite{DV99,DV04,LN00,CDP94,SW90},
or during any window of first $t$ steps \cite{GP98}, requiring that after some finite time there is a
long enough fault-free computation \cite{D74, D00} etc. 
While the probabilistic model is analyzed with respect to the expected behavior,
the deterministic models have been mostly analyzed for the worst case scenario.

We shall focus our attention on synchronous point-to-point distributed systems, i.e. systems
in which the communication is performed by sending messages along links connecting pairs of vertices.
Moreover, the vertices are synchronized by a common clock, and the delivery of every message
takes exactly one time unit. This model has been widely considered
\cite{CDP94,DV99,DV04,LN00,GP98,KKR03,DKKS06,CDP91,P91,PP05,SW90} not only for its 
theoretical appeal, but for its practical relevance as well (e.g. many wireless networking
standards, like IEEE 802.11, or GSM, operate in discrete time steps). We shall consider
only one type of failures: {\em message loss}. 

The oldest deterministic model of faults considered in this setting is the static
model \cite{AGHK96,BH86}, in which it is assumed 
that at most a fixed constant number $k$ of messages may be lost in every
step, and moreover, the failures are always located
on the same links. Later, other models have been considered, too, like the dynamic
model \cite{DV99,DV04,LN00,CDP94,SW90} in which the $k$ failures may be
located on arbitrary links in every step, linearly bounded faults \cite{GP98},
fractional faults \cite{KKR03}, etc.

We continue in the analysis of the {\em fractional model with threshold} from
\cite{DKKS06}. Here, the number of messages lost in one time step is
bounded by the maximum of a fixed threshold $T$ and a fixed fraction $\alpha$
of sent messages. This restriction implies that if, in a given step, fewer than $T$ messages
are sent they may all be lost. On the other hand,
if there are many messages sent, at least a fixed fraction
$1-\alpha$ of them is delivered. The threshold $T$ is always assumed to be
one less than the edge connectivity,
since this is the largest value under which the network stays connected. This model has been developed
in order to avoid some unrealistic special cases of static and dynamic models (the number of faults
is independent on the actual network traffic), as well as those of fractional model (if just one
message is being sent, its delivery is always guaranteed).

The broadcasting problem is a crucial communication task in the study of distributed systems 
(e.g. \cite{HHL88}).
One vertex, called initiator, has a piece of information that has to be distributed among
all remaining vertices. The broadcasting has not only been used as a test-bed application for the study
of the complexity of communication in various communication models, but has served as a 
building stone of many applications (e.g. \cite{SAA98}) as well.

We analyze the broadcasting in complete graphs and hypercubes. The broadcasting time in these graphs has been 
studied in the static \cite{F92}, 
dynamic \cite{DV04,DV99,LN00}, and simple threshold \cite{DKKS06}\footnote{
If the number of messages sent in a given time step is less than the edge
connectivity $c(G)$ in the simple threshold model, all of them may be lost.
Otherwise at least one of them is delivered.
} 
models, and the results are summarized in Table~\ref{tab:results}.

\begin{table}[htb]
\label{tab:results}
{\small
\begin{center}\begin{tabular}{|l|l|l|l|}\hline
Model & $K_n$, chordal & $K_n$ & $Q_d$, $n=2^d$ \\
      & sense of direction & unoriented &  \\\hline
static &  $\Theta(1)$ & $\Theta(1)$ & $d+1$ \cite{F92}\\
dynamic & $\Theta(1)$ & $\Theta(1)$ \cite{LN00}& $d+2$ \cite{DV99} \\
fractional & $\Theta(\log n)$ & $\Theta(\log n)$ \cite{KKR03} & $O(d^3)$ \cite{KKR03}\\
simple threshold & $\Omega(n)$, $O(n^2)$ \cite{DKKS06}& $\Omega(n^2)$, $O(n^3)$ \cite{DKKS06} & $O(n^4d^2)$ \cite{DKKS06}
\\\hline
\end{tabular}\end{center}
}
\caption{Known time complexities of the complete broadcasting in various models.}
\end{table}

\begin{table}[htb]
\label{tab:results-fraction-threshold}
{\small
\begin{center}\begin{tabular}{|l|l|l|}\hline
Scenario & Almost complete & Complete \\
         & broadcasting    & broadcasting \\
\hline
$K_n$, unoriented & $O(\log n)$ & $\Omega(\log n)$ \cite{KKR03}, $O(n^3)$ \cite{DKKS06}\\
$K_n$, chordal sense of direction & $O(\log n)$ & $\Omega(\log n)$ \cite{KKR03}, $O(\log n)$ \\
$K_n$, $\alpha<0.55$ & $O(\log n)$ & $\Omega(\log n)$ \cite{KKR03}, $O(\log n)$ \\
$Q_d$ & $O(d^2)$ & $\Omega(d)$, $O(n^4d^2)$ \cite{DKKS06} \\
\hline
\end{tabular}\end{center}
}
\caption{Results for the complete and almost complete broadcasting in the
fractional model with threshold.}
\end{table}

We address a natural relaxation of the broadcasting problem in which we allow
a small  constant number of vertices to stay uninformed in the end (a problem called
{\em almost complete broadcasting}), and analyze the worst case time needed to solve
the problem. Our main motivation to study almost complete broadcasts is the fact that in 
large faulty networks it is often
vital to finish a communication task fast, even subject to some small error. In the
probabilistic setting, this is modelled by allowing a failure probability 
that tends to zero with increasing network size: in the worst case the task is not successful but this
worst case scenario has a small probability. 
Since in our deterministic setting we study the worst case, another
model of allowed error must be chosen. If we look at the broadcast as an optimization problem
where the task is to inform as many vertices as possible, it is natural to introduce
a constant additive error by allowing a constant number of vertices to stay uninformed\footnote{so that the 
uninformed vertices comprise at most an $O(1/n)$ fraction
of all vertices}.

For complete graphs and hypercubes, we show that the problem can be solved
in time $O(D\log n)$, where $D$ is the diameter of the graph and $n$ is the number of its nodes.

Moreover, we show that if the complete graph is equipped with the chordal sense of direction, complete
broadcasting can be performed in time $O(\log n)$. This is asymptotically optimal since the broadcasting time in 
the fractional model is a lower bound for the fractional model with
threshold. Similarly we show that the broadcasting can be completed in time
$O(\log n)$ for values $\alpha<0.55$. The overview of the results can be found in
Table~\ref{tab:results-fraction-threshold}.

\section{Definitions}

We consider a synchronous, point-to-point distributed system with a coordinated start-up.
The system consists of a number of nodes and a number of communication links connecting some pairs of nodes.
The system is modelled by an undirected graph, in which vertices correspond to nodes and edges correspond
to communication links. In this respect, we shall use the terms ``node'' and ``vertex'' interchangeably. 
Sometimes we need to argue about outgoing and incoming links; in this cases we consider a directed
graph obtained from the undirected one by replacing each edge by two opposite arcs. 

At the beginning of the computation all nodes are active and start performing the given protocol.
The computation consists of a number of steps: at the beginning of each step, messages sent during
the previous step are delivered to their destinations, then each vertex performs some
local computation, possibly sending some messages\footnote{
i.e. a vertex may send different message to each of its neighbors in one step}, and the next step begins.

The failure model we consider is the {\em fractional dynamic
faults with threshold} from \cite{DKKS06}, which can be described as a game between the
algorithm and an adversary: in a time step $t$ the algorithm sends
$m_t$ messages and the adversary may destroy up to 
$$F(m_t) = \  \max\{c(G)-1,\lfloor\alpha\ m_t \rfloor\}$$
of them, where $c(G)$ is the edge connectivity of the
graph and $\alpha$ is a known, fixed constant $0<\alpha<1$. There is no built-in mechanism 
of acknowledgements, so the sender node is not informed whether a particular message was delivered
or destroyed.

We consider the problem of broadcasting, where an initiator has a piece of information to
be transmitted to all remaining vertices. We call a broadcast {\em complete} if all vertices have the information after
the termination of the algorithm. A broadcast is called {\em almost-complete}
if there is a fixed constant $c$ (independent on the network size) 
such that after the termination there are at most $c$ uninformed
vertices. Hence, to prove the existence of an almost-complete broadcasting algorithm for
a family of graphs ${\cal G}$, one has to prove
that there exists a constant $c$ such that for each $G\in{\cal G}$ the broadcasting algorithm
informs all but $c$ vertices of $G$.

In all presented algorithms only the informed vertices send
messages. Arcs (i.e. directed edges)
leading from an informed vertex can be classified as being either active, 
passive or hyperactive during the computation:

\begin{definition}
Let $e$ be an arc leading from an informed vertex. We call $e$ {\em active} if it leads to
an uninformed vertex.
We call an arc $e$ {\em passive}, if some message has been delivered via the
opposite arc of $e$. 
Finally, we call an arc $e$ {\em hyperactive} if it leads to an informed vertex, and is not passive.
\end{definition}

If the arc $e$ is passive, the source vertex of $e$ is aware of the
fact that the destination vertex of $e$ has already been informed.
The main idea of our algorithms is to perform appropriate number of {\em
simple rounds} defined as follows:

\begin{definition}
A {\em simple round} consists of two time steps.  In the first step, every
informed vertex sends a message along each of its incident arcs, excluding the passive
ones.\footnote{In this step, a message is sent via all active and
hyperactive arcs. The former can inform new vertices, the latter exhibit
only useless activity. However, the algorithm can not distinguish between
active and hyperactive arcs.} In the second step, all vertices that have
received a message send an acknowledgement (and mark the arc as passive).
Vertices that receive acknowledgement mark the corresponding arc as
passive.
\end{definition}

For the remainder of this paper, let $0<\alpha<1$ be a known fixed constant, and
let us denote $$X:=\frac{1}{\alpha(1-\alpha)}$$

The rest of the paper is organized as follows. In the next two sections we present algorithms for
the almost-complete broadcasting on complete graphs and hypercubes, respectively, that run in
time $O(D\log n)$. Then we show how
to obtain broadcast in complete graphs equipped with chordal sense of direction, and for unoriented complete
graphs for $\alpha<0.55$, having the same time complexity.

Some technical parts have been omitted from this paper, and can be found
in the appendix.

\section{Complete Graphs}
\label{sec:complete}

In a complete graph $K_n$, all $n$ vertices have degree $n-1$, and $n-1$ is also the edge connectivity.
Hence, in each step $t$ the adversary can destroy up to $\max\{n-2,\lfloor\alpha m_t\rfloor\}$ messages, where $m_t$ is
the number of messages sent in the step $t$. In this section we present an algorithm that
informs all but a constant number of vertices in logarithmic time. The idea of the algorithm
is very straightforward -- just repeat simple rounds sufficiently many times. However, the arguments
given in the analysis of a simple round below hold only if there are enough informed vertices participating
in the round. To satisfy this requirement two steps of a simple greedy algorithm are performed, during
which each informed vertex just sends the message to all vertices. After two steps of this algorithm,
the number of informed vertices is as shown in Lemma~\ref{lm:grd}.

\begin{lemma}
\label{lm:grd}
After two steps of the greedy algorithm, at least 
$$1+\min\left\{\frac{n}{2},(n-1)(1-\alpha)\right\}$$
vertices are informed.
\end{lemma}

After these two steps, the algorithm performs a logarithmic number of simple rounds. To show that logarithmic
number of simple rounds is sufficient to inform all but one vertex we first
provide a lower bound on  the number of 
acknowledgements delivered in each round, and then we show that each delivered acknowledgement
decreases a certain measure function.

\begin{theorem}
\label{thm:AlmostBroadcast-Kn}
Let $\varepsilon>1$ be an arbitrary constant.
For large enough $n$ it is possible to inform all but at most $X\varepsilon$ vertices in logarithmic time.
Moreover, the number of remaining hyperactive arcs is at most $X(n-2)$.
\end{theorem}
\begin{proof}
At the beginning, two steps of the greedy algorithm are executed.  Then, a
logarithmic number of simple rounds is performed.
Now consider the situation at the beginning of the $i$-th round. Let $k_i$ 
be the number of uninformed vertices, and $h_i$ the number
of hyperactive arcs. 
We claim that if $k_i>X\varepsilon$ or $h_i>X(n-2)$ then at least $\left[k_i(n-k_i)+h_i\right](1-\alpha)^2$ 
acknowledgements
are delivered in this round. Since there are $k_i(n-k_i)+h_i$ messages sent in this round, in order to prove the claim
it is sufficient to show that $\alpha(1-\alpha)\left[k_i(n-k_i)+h_i\right]\ge n-2$. 
Obviously, if $h_i>X(n-2)$ the inequality holds,
so consider the case $k_i>X\varepsilon$. We prove that in this case $k_i(n-k_i)\ge X(n-2)$, i.e. $k_i^2-nk_i+X(n-2)\le0$.
Let $f(n):=1/2\left(n-\sqrt{n^2-4X(n-2)}\right)$; the roots\footnote{Assume
that $n$ is large enough such that $f(n)$ is real number.} of the equation
$k_i^2-nk_i+X(n-2)=0$ are $f(n)$ and $n-f(n)$, so we want to show that $f(n)\le
k_i\le n-f(n)$. 
Since $\lim_{n\mapsto\infty}f(n)=X$, we get that $k_i>X\varepsilon>f(n)$ holds for large enough $n$. Hence, the
only remaining step is to show the inequality $k_i\le n-f(n)$. From Lemma~\ref{lm:grd} it follows that 
$n-k_i>\min\left\{n/2,(n-1)(1-\alpha)\right\}$. Since $f(n)<n/2$, if $n-k_i>n/2$ it holds $k_i<n-f(n)$. 
So let us suppose
that $n-k_i>(n-1)(1-\alpha)$, i.e. $k_i<1+\alpha(n-1)$. Let 
$n\ge\frac{\varepsilon+\alpha(1-\alpha)^2}{\alpha(1-\alpha)^2}$.
Then it holds for large enough $n$ that
$$k_i< 1+\alpha n-\alpha\le n-\frac{\varepsilon}{\alpha(1-\alpha)}=n-\varepsilon X\le n-f(n).$$

We have proved that if $k_i>X\varepsilon$ or $h_i>X(n-2)$ then at least 
$$\left[k_i(n-k_i)+h_i\right](1-\alpha)^2$$ 
acknowledgements
are delivered in round $i$.

To conclude the proof we show that after logarithmic number of iterations we get $k_i\le X\varepsilon$ and 
$h_i\le X(n-2)$.
Let $M_i:=2(n-1)k_i+h_i$; then every delivered acknowledgement decreases $M_i$ 
by at least one: indeed, if the acknowledgement
was delivered over a hyperactive arc, $h_i$ decreases by 1. If, on the other hand, the acknowledgement was delivered over an
active arc, the number of uninformed vertices is decreased by at least
one, and the number of hyperactive arcs is increased by
at most $2n-3$ (new hyperactive arcs are between the newly informed vertex and any other vertex, with the exception of the
arc that delivered the acknowledgement which is passive).

From Lemma~\ref{lm:grd} it follows that either $n-k_i>n/2$ or $n-k_i>(n-1)(1-\alpha)$. In the first case it follows
that at least $(1-\alpha)^2\left[k_i(n-k_i)+h_i\right]>(1-\alpha)^2\left[k_in/2+h_i\right]\ge\frac{(1-\alpha)^2}{4}M_i$
acknowledgements are delivered. In the second case we get that at least
$(1-\alpha)^2\left[k_i(n-k_i)+h_i\right]>(1-\alpha)^2\left[k_i(n-1)(1-\alpha)+h_i\right]\ge\frac{(1-\alpha)^3}{2}M_i$
acknowledgements are delivered.
Let $c:=\min\{\frac{(1-\alpha)^2}{4},\frac{(1-\alpha)^3}{2}\}$, then obviously every iteration decreases the
value of $M_i$ at least by factor $c$. Since the value of $M$ at the
beginning of the algorithm is $M_1=O(n^2)$, $\log_{1/c}M_1=O(\log
n)$ steps are sufficient to inform all but a constant number (at most
$X\varepsilon$) of vertices and to ensure that the number of remaining
hyperactive arcs is linear (at most $X(n-2)$).
\end{proof}

\section{Hypercubes}

In this section we consider $d$-dimensional hypercubes. The hypercube $Q_d$
has $2^d$ vertices, and both diameter and edge connectivity are $d$. 
We present an algorithm that
informs all but a constant number of vertices in time $O(d^2)$.

The general idea is the same as for complete graphs: first we perform two initialization steps
to make sure there are enough informed vertices for the subsequent analysis to hold. Next, simple rounds
are repeated for a sufficient number of times. The analysis, however, is more complicated in this case.

The next lemma covers the initialization steps. In the first step, the initiator sends a message to all
its neighbors, and at least one of these messages is delivered. In the second step, the initiator sends
a message to all its neighbors again; moreover, each of the vertices informed in the first step sends a message
to all its neighbors except the initiator. 

\begin{lemma}
\label{lm:Q:firststeps}
After the first two steps of the algorithm, at least $\frac{1-\alpha}{2}(2d-1)$ vertices
are informed.
\end{lemma}

\noindent
For the rest of this section we suppose that there are at least $\frac{1-\alpha}{2}(2d-1)$ informed vertices.
We show that after $O(d^2)$ simple rounds all but some constant number of vertices are informed, 
and there are only
linearly many hyperactive arcs. At the end of this section, we shall be able to prove the
following theorem.

\begin{theorem}
\label{thm:hypercube}
Let  $\varepsilon\in(0,1)$ be an arbitrary constant. For large enough $d$ it is possible
to inform all but at most $X/(1-\varepsilon)$ vertices of $Q_d$ within $O(d^2)$ time steps. Moreover,
the number of remaining hyperactive arcs is at most $X(d-1)$.
\end{theorem}

In our analysis we need to assert that enough acknowledgements are delivered, given the number
of informed vertices. To bound the number of sent messages, we 
rely heavily upon the following isoperimetric inequality due to Chung et.\ al.\cite{C+88}:

\begin{claim}{{\rm\bfseries\cite{C+88}}}
\label{HypercubeBorder}
Let $S$ be a subset of vertices of $Q_d$. The size
of the edge boundary of $S$, denoted as $\partial(S)$ is defined as the
number of edges connecting $S$ to $Q_d\setminus S$.
Let $\partial(k)=\min_{|S|=k}\partial(S)$, and let $\lg$ denote the logarithm of base 2.
It holds that 
$$\partial(k)\ge k(d-\lg k)$$
\end{claim}

The first step in the analysis is to prove that if there are enough uninformed vertices, or enough
hyperactive arcs at the beginning of a round $i$, then sufficiently many acknowledgements
are delivered in this round:

\begin{lemma}
\label{lm:HypercubePass}
Consider a $d$-dimensional hypercube with $k$ non-informed vertices and
$h$ hyperactive arcs. Let  $\varepsilon\in(0,1)$ be an arbitrary constant, and let $k>X/(1-\varepsilon)$ or $h>X(d-1)$. 
Then in the second step of a simple
round at least $\beta(h+\partial(k))$ acknowledgements are delivered,
where $\beta=(1-\alpha)^2$.
\end{lemma}
\begin{proofsk}
Let $S$ be the set of informed vertices. In the first step of the round, $h+\partial(S)$ messages are sent.
Since the edge boundary of informed and uninformed vertices is the same,
at least $h+\partial(k)$ messages are sent
in the first step of the round.
The idea of the proof is to show that $\alpha(h+\partial(k))\ge d-1$, 
so in the first step at most $\alpha(h+\partial(k))$ messages are lost,
and at least $(1-\alpha)(h+\partial(k))$ of them are delivered. Next we prove that 
$\alpha(1-\alpha)(h+\partial(k))\ge d-1$,
so in the second step  at least $(1-\alpha)^2(h+\partial(k))$ messages are delivered. Since $1-\alpha<1$, it is sufficient
to prove that $\alpha(1-\alpha)(h+\partial(k))\ge d-1$.
If $h>X(d-1)$ then clearly $h+\partial(k)\ge X(d-1)$ and the statement holds. 
Hence, the main goal of the proof is to show that for $k>X/(1-\varepsilon)$,
it holds $\partial(k)\ge X(d-1)$. To do so, the inequality
$2^d-k\ge \frac{1-\alpha}{2}(2d-1)$, which is granted by Lemma
\ref{lm:Q:firststeps}, is used.
\end{proofsk}

In the rest of the proof of Theorem~\ref{thm:hypercube} we show that $O(d^2)$ simple rounds are sufficient to
inform almost all vertices. The analysis is divided into two parts. In the first part we prove that within $O(d^2)$
rounds at least $2^d/3$ vertices are informed. In the second part we show that another $O(d^2)$ rounds
are sufficient to finish the algorithm.

\begin{lemma}
\label{lm:Q:part:I}
After performing $O(d^2)$ simple rounds on $Q_d$ at
least $2^d/3$ vertices are informed.
\end{lemma}
\begin{proofsk}
Let $l:=2^d-k$ be the number of informed vertices. 
From Lemma~\ref{lm:HypercubePass} it follows that at least $\beta \partial(k)$ 
acknowledgements are delivered in one simple round. Since the edge boundary of informed
vertices is also the boundary of uninformed vertices, the number of delivered acknowledgements
is at least $\beta \partial(l)$. Furthermore, every delivered acknowledgement adds one passive arc,
so the number of passive arcs grows at least by $\beta \partial(l)$ each round, which 
we show to be at least a factor of $\left(1+\frac{1}{\frac{d}{\beta\lg 3}}\right)$.
Because the number of passive arcs cannot grow over $d2^d/3$ without informing at least $2^d/3$ 
vertices, we get the statement of the lemma.
\end{proofsk}

\begin{lemma}
\label{lm:Q:part:II}
Let $\varepsilon\in(0,1)$ be an arbitrary constant, and let
$k_i\le(2/3)2^d$ be the number of uninformed vertices and $h_i$ the number
of hyperactive arcs of an $d$-dimensional hypercube at the beginning of round $i$. 
Then after $O(d^2)$ simple
rounds there are at most
$X/(1-\varepsilon)$ uninformed vertices and at most $X(d-1)$ hyperactive arcs.
\end{lemma}
\begin{proofsk}
Similarly to the proof of Theorem \ref{thm:AlmostBroadcast-Kn}, let us
consider the measure $M_i:=2dk_i+h_i$ which decreases
with every acknowledgement delivered. We show that as long as the
requirements of Lemma~\ref{lm:HypercubePass} hold, $M_i$ decreases
in each round by a factor  $\left(1+\frac{\beta\lg (2/3)}{d}\right)$.
Since $M_i\le(5/3)d2^d$, we get the statement of the lemma.
\end{proofsk}

Combining Lemma~\ref{lm:Q:firststeps} with Lemma~\ref{lm:Q:part:I} and Lemma~\ref{lm:Q:part:II}
completes the proof of Theorem~\ref{thm:hypercube}.

\section{Complete Broadcast in Complete Graphs}

In Section~\ref{sec:complete} we have shown how to inform all but some constant number
of vertices in a complete graph $K_n$ in time $O(\log n)$. A natural question
is to ask if it is possible to inform also the remaining vertices in the same
time complexity. In this section we partially answer this question. In particular,
we show in the following subsection 
that if the graph is equipped with a chordal sense of direction, then the 
complete broadcasting can be performed in time $O(\log n)$. In the subsequent
subsection, we show that if the constant $\alpha<0.55$, complete broadcast can be
performed in time $O(\log n)$ without the sense of direction, too.

\subsection{Chordal Sense of Direction}

\begin{figure}[htb]
\centerline{\includegraphics[width=4cm]{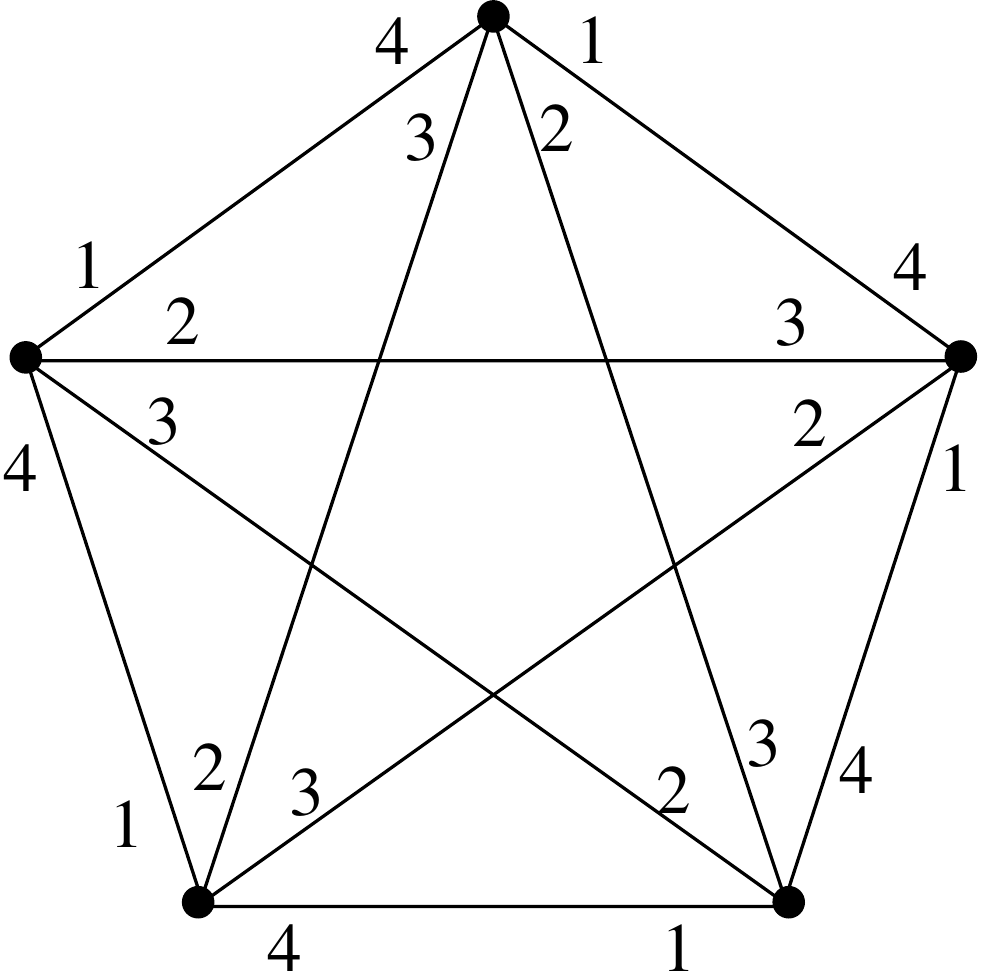}}
\caption{$K_5$ with the chordal sense of direction}
\label{fig:sod}
\end{figure}

Let us consider a complete graph with a fixed Hamiltonian cycle $\cal C$ (unknown to the vertices).
We say that the complete graph has a chordal sense of direction if in every
vertex the incident arcs are labeled by the clockwise distance on $\cal C$ (see 
Figure~\ref{fig:sod}). The notion of a sense of direction has been defined
formally for general  graphs, and it has been known to significantly
reduce the complexity of many distributed tasks (e.g. \cite{F+97,F+98}).

We show how to perform a complete broadcast on
a complete graph with the sense of direction in time $O(\log n)$. The process consists of three steps.
First, using Theorem~\ref{thm:AlmostBroadcast-Kn}, all but a constant
number of vertices are informed. In the second phase the information is delivered to all but one
vertex. In the last phase the remaining single vertex is informed.

The sense of direction is essential to our algorithm. Since there
is a unique initiator of the broadcasting, all vertices can derive unique
identifiers defined as their distance on $\cal C$ from the initiator. Furthermore, the
sense of direction allows each vertex to know the identifier of a
destination vertex of any of its incident arcs. 

\begin{lemma}
\label{lm:AllButOne-Kn}
It is possible to inform all vertices but one on complete graphs with chordal
sense of direction in time $O(\log n)$. Furthermore, after finishing the
algorithm vertex $0$ or vertex $1$ knows a constant number of candidates
for the uninformed vertex.
\end{lemma}
\begin{proof}
The outline of the algorithm is as follows: At first the algorithm from
Theorem \ref{thm:AlmostBroadcast-Kn} is performed, which ensures that all but
a constant number of vertices are informed. Afterwards a significant group
of vertices negotiate a common set $U$ of {\em candidates} for uninformed
vertices, such that all uninformed vertices are in $U$ and the size of $U$
is constant. The vertices then cooperate to inform all vertices in $U$ but
one. As a side effect, the set $U$ will be known to vertex $0$ or vertex
$1$, hence satisfying the second claim of the lemma.
Now we present this algorithm in more detail:

\begin{description}
\item[Phase 1] Run the algorithm from Theorem \ref{thm:AlmostBroadcast-Kn}.
This phase takes $O(\log n)$ time and ensures that there are at most
$X\varepsilon$ uninformed vertices and at most $X(n-2)$ hyperactive arcs.

\item[Phase 2] Each vertex $v$ that has at most $3X(1+\varepsilon)$ non-passive
(i.e.\ active or hyperactive) links leading to the set of vertices
$U_v$ sends a message containing $U_v$ to vertices with number $0$ and $1$.

Now we show that at least one of these messages is delivered. It is easy to see
that there are at least $2n/3$ vertices satisfying the above-mentioned condition,
otherwise there would be more than $n/3$ vertices with at least
$3X(1+\varepsilon)$ non-passive links, so there would be more than
$nX(1+\varepsilon)$ active or hyperactive arcs. But since the number of
uninformed vertices is at most $k\le X\varepsilon\le n/2$ for large $n$, there
are $k(n-k)\le X\varepsilon(n-X\varepsilon)$ active arcs. So the total number of
active or hyperactive arcs is at most $X\varepsilon(n-X\varepsilon)+X(n-2)\le
Xn(1+\varepsilon)$, which is a contradiction.

The rest of the algorithm will be time-multiplexed into two parts. In even time
steps, the case that the vertex $0$ received a message in phase 2 is
processed. In odd time steps, the case that the vertex $1$ received a
message is processed analogously. Hence, we can restrict to the first case
in the rest of the algorithm description. As there are only two cases the
asymptotic complexity of the algorithm is unaffected by the multiplexing.

\item[Phase 3] The vertex $0$ received at least one message containing a
set of possibly uninformed vertices. It is obvious that the set of
uninformed vertices is a subset of every received message. Hence the
set $U$ can be defined as the intersection of the received messages:
Indeed, every uninformed vertex is in $U$ and the size of $U$ is at most
$3X(1+\varepsilon)=O(1)$. The set $U$ is then distributed using the algorithm in
Theorem \ref{thm:AlmostBroadcast-Kn} among at least $n-X\varepsilon$ vertices in
time $O(\log n)$.

\item[Phase 4] There are at least $n-X\varepsilon$ vertices aware of the set
$U$. In this phase they cooperate to inform all but one vertex in $U$,
using an idea similar to Lemma 2 in \cite{DKKS06}: every vertex aware of
the set $U$ iterates through all pairs $[i,j]$ ($i,j\in U$) in
lexicographical order; in each time step it sends the original message
to both vertices $i$ and $j$. Since in each time step at least $2n-X\varepsilon$
messages are sent, at least one of them is delivered (for large enough
$n$). As all vertices process the same pair $[i,j]$ in every time step,
this ensures that a new vertex is informed whenever both $i$ and $j$ were
uninformed. Hence, at the end of this phase all vertices but one are
informed. The time complexity of this phase is $O(|U|^2)=O(1)$.
\end{description}

It is obvious that after finishing the Phase 4 the claim of the Lemma
holds.
\end{proof}

Finally, we show how to inform the last remaining vertex, thus proving
the following theorem:

\begin{theorem}
\label{thm:K:sod}
It is possible to perform broadcasting on complete graphs with chordal
sense of direction in time $O(\log n)$.
\end{theorem}
\begin{proofsk}
Suppose that after performing the algorithm from Lemma \ref{lm:AllButOne-Kn}
all vertices with the exception of some vertex $v$ are informed and 
vertex $0$  knows a set $U$ of constant size containing
candidates for $v$. The algorithm from Lemma \ref{lm:AllButOne-Kn}
is used again to broadcast $U$ with two possible outcomes: either 
$v$ was informed during the broadcast, or all other vertices have the same set of 
candidates, which they try to inform one by one.
\end{proofsk}

\subsection{Without Sense of Direction}

As a last result in this paper we show that it is possible to perform broadcasting on 
complete graphs 
in time $O(\log n)$ for small values of $\alpha$ (i.e. $\alpha\lesssim
0.55$) even without the sense of direction. The idea is to use the 
algorithm from Theorem \ref{thm:AlmostBroadcast-Kn} to inform all but constantly many
vertices. Next, instead of repeating 2-step simple rounds, some $\log n$-step extended rounds
are repeated, such that each extended round informs a yet uninformed vertex. During an extended round
messages are sent for $O(\log n)$ steps in such a way that 
in every step the number of hyperactive arcs
is decreased by some factor\footnote{in this part we need the assumption
that $\alpha$ is small enough} unless a new vertex is informed.

\begin{theorem}
\label{thm:K:nosod}
Let $1-\alpha-2\alpha^2+\alpha^3>0$. Then it is possible to perform
broadcasting on complete graphs without sense of direction in time $O(\log
n)$.
\end{theorem}
\begin{proof}
The algorithm is described as Algorithm \ref{alg:complete}.

\begin{algorithm}
\caption{Complete graphs without sense of direction}
\label{alg:complete}
\begin{algorithmic}[1]
\State perform almost-complete broadcast according to Theorem \ref{thm:AlmostBroadcast-Kn}
\State let $k$ denote the number of uninformed vertices, let $h$ denote the number of hyperactive arcs
\Loop{} $L_1$ times  \label{alg:complete-L1}
  \Comment{Perform $L_1$ extended rounds}
  \Loop{} $L_2(n)$ times  \label{alg:complete-L2}
    \Comment{In each iteration $h$ decreases by a constant factor}
    \State $E:=\mbox{set of all currently active or hyperactive arcs}$; $P:=\emptyset$
    \label{alg:complete-set}
    \Loop{} $L_3$ times  \label{alg:complete-L3}
      \State send the message via all arcs in $E\cup P$ \label{alg:complete-send} 
      \State $P:=P\cup \left\{e \mid \begin{array}{l}\mbox{a message has been
                  delivered in this step}\\\mbox{via the opposite arc of $e$}\end{array}\right\}$ 
             \label{alg:complete-receive} 
    \EndLoop  \label{alg:complete-L3E}
  \EndLoop  \label{alg:complete-L2E}
  \Loop{} $L_4(n)$ times  \label{alg:complete-L4}
    \Comment{Inform new vertex and decrease $a$}
    \State perform one simple round
  \EndLoop  \label{alg:complete-L4E}
\EndLoop  \label{alg:complete-L1E}
\end{algorithmic}
The values of $L_1$, $L_2(n)$, $L_3$ and $L_4(n)$ are specified in the
analysis of the algorithm, such that $L_1,L_3=O(1)$ and $L_2(n),
L_4(n)=O(\log n)$.
\end{algorithm}

At first, the algorithm from Theorem \ref{thm:AlmostBroadcast-Kn} is performed,
ensuring that there are at most $k\le X\varepsilon$ uninformed vertices
and at most $h\le X(n-2)$ hyperactive arcs ($X$ and $\varepsilon$ have the
same meaning as in Theorem \ref{thm:AlmostBroadcast-Kn}). The purpose of one
iteration of the loop on lines \ref{alg:complete-L1}--\ref{alg:complete-L1E} is to
inform at least one uninformed vertex. Taking $L_1:=X\varepsilon=O(1)$
ensures that all vertices will be informed.

The loop on lines \ref{alg:complete-L2}--\ref{alg:complete-L2E} reduces
the number of hyperactive arcs to zero unless a new vertex is informed. One
iteration of this loop either informs a new vertex or reduces the number of
hyperactive arcs from $h$ to $(1-Y/2)h$, where $0<Y<1$ is a constant
(depending on $\alpha$) defined later. Hence the number of hyperactive arcs
decreases exponentially with number of iterations of the loop and
$\log_{1/(1-Y/2)}h$ iterations are sufficient to eliminate all hyperactive arcs.
Since the condition $h\le X(n-2)$ holds before every execution of the loop
(this is provided either directly by Theorem \ref{thm:AlmostBroadcast-Kn}  or
by the loop on lines \ref{alg:complete-L4}--\ref{alg:complete-L4E}), we can
define $L_2:=\log_{1/(1-Y/2)}(X(n-2))=O(\log n)$.

Now we describe one iteration of the loop on lines
\ref{alg:complete-L2}--\ref{alg:complete-L2E}. 
We distinguish two types of arcs that are hyperactive at the beginning of
the considered iteration: An arc $e$ is a {\em single hyperactive
arc} if and only it is hyperactive and the opposite arc of $e$ is passive
at the beginning of the iteration. Otherwise (i.e.\
if both $e$ and the opposite arc of $e$ are hyperactive at the beginning of
the iteration), $e$ is a {\em double hyperactive arc}.

Let $E$ be the set of all active or hyperactive arcs at the beginning of
the iteration, and $P$ be the set of all arcs opposite to arcs through which
some message has been delivered in the current iteration. Furthermore, let
$k'$ be the number of uninformed vertices at the beginning of the current
iteration, $h'$ be the number of hyperactive arcs at the beginning of the
current iteration and $p=|P\setminus E|$ be number of arcs in $P$ that were passive at the
beginning of the current iteration. It clearly
holds that $|E|=k'(n-k')+h'$ and that $k'(n-k')+h'+p$ messages
are sent on every execution of line \ref{alg:complete-send}. Since at least
$n-1$ messages are lost (because we may assume that no new vertex is
informed), at most $\alpha(k'(n-k')+h'+p)$ of them are lost, i.e. at
least $(1-\alpha)(k'(n-k')+h'+p)$ are delivered. 

Now assume by contradiction that the number of hyperactive arcs does not
decrease below $(1-Y/2)h'$, and no new vertices are informed during the
current iteration of the loop on lines
\ref{alg:complete-L2}--\ref{alg:complete-L2E}. Consider any message
delivered over an arc $e$ which is a double hyperactive arc or an arc in
$P\setminus E$; it is easy to see that the opposite arc of $e$ is passive after the
delivery and that it was hyperactive at the beginning of the iteration. This fact yields that at most $(Y/2)h'$ messages are
delivered over a double hyperactive arc or an arc in $P\setminus E$ on any execution
of line \ref{alg:complete-send}.

Now we show a lower bound on the number of messages that pass over double
hyperactive arcs or arcs in $P\setminus E$ or single hyperactive arcs whose opposite
arcs are not in $P\setminus E$. Intuitively, every such message ensures some progress of the
algorithm, since either an arc is made passive (in the first two cases) or
a new arc is added to $P\setminus E$ (in the third case). As no messages
passes over active arcs by our assumption, and at most $p$ messages pass
over single hyperactive arcs whose opposite arcs are in $P\setminus E$, there are at
least $(1-\alpha)(k'(n-k')+h'+p)-p$ messages satisfying one of these
three cases. Using the inequalities $k'(n-k')\ge n-1$ and $p\le h'$
yields $(1-\alpha)(k'(n-k')+h'+p)-p\ge (1-\alpha)(n-2)+(1-2\alpha)h'$.
Because $h'\le X(n-2)$ which is equivalent to $(n-2)\ge
\alpha(1-\alpha)h'$,
we have $(1-\alpha)(k'(n-k')+h'+p)-p\ge
(1-\alpha-2\alpha^2+\alpha^3)h'$. Defining $Y:=1-\alpha-2\alpha^2+\alpha^3$,
which is positive and less than one by the assumption of the Lemma, we have shown
that there are at least $Yh'$ messages satisfying one of the three cases.

However, at most $(Y/2)h'$ of them satisfies the first two
cases, hence there are at least $(Y/2)h'$ arcs added to $P$ in
every execution of line \ref{alg:complete-receive}. So taking
$L_3:=2/Y+1$ ensures that $P$ contains opposite arcs to all single hyperactive arcs at the
beginning of the last iteration of the loop on lines
\ref{alg:complete-L3}--\ref{alg:complete-L3E}. However, this is a
contradiction with the fact that new arcs are added to $P$ at line
\ref{alg:complete-receive}.

We conclude the proof with the analysis of the loop on lines
\ref{alg:complete-L4}--\ref{alg:complete-L4E}. In the first iteration of
the loop a new vertex is informed, because there are no hyperactive arcs
left after the loop on lines \ref{alg:complete-L2}--\ref{alg:complete-L2E}
finished (unless the new vertex has already been informed in that loop).
Due to Theorem \ref{thm:AlmostBroadcast-Kn}, next $O(\log n)$ iterations
are sufficient to ensure that $h\le X(n-2)$, which is an invariant required
by the loop on lines \ref{alg:complete-L2}--\ref{alg:complete-L2E}. Hence
putting $L_4(n):=O(\log n)$ (according to Theorem
\ref{thm:AlmostBroadcast-Kn}) is sufficient to make the algorithm work
correctly in time $L_1(L_2(n)L_3+L_4(n))=O(\log n)$.
\end{proof}

\section{Conclusions, Open Problems, and Further Research}

We have studied the problem of almost complete broadcast under the model of 
fractional dynamic faults with threshold. We showed that both in complete graphs
and in hypercubes, it is possible to inform all but constantly many vertices in time
$O(D\log n)$ where $D$ is the diameter of the graph and $n$ is the number of vertices.

Moreover, we have proved that if the complete graph is equipped with the chordal sense of direction,
or the parameter $\alpha<0.55$, a complete broadcast can be performed in time $O(\log n)$.

This research leaves many open questions and directions for further research, from which
we mention at least a few. One obvious
question is to ask if it is possible to perform a complete broadcast in
complete graphs also for large values of $\alpha$ in polylogarithmic time.
The difficulty of broadcast in the fractional dynamic model with
threshold stems from the fact that, in order to inform the last few vertices, all informed
vertices must cooperate very tightly.
In general, the relationship between the almost complete and complete
broadcast in various models is worth studying. We have also not considered non-constant
values of $\alpha$. It would be interesting to extend our results to more general
classes of graphs.

We finish by noting that there is a lack of any non-trivial lower bounds in the
model of fractional faults with threshold. 

{\bibliography{references}}

\begin{thebibliography}{10}

\bibitem{AGHK96}
R.~Ahlswede, L.~Gargano, H.~S. Haroutunian, and L.~H. Khachatrian.
\newblock Fault-tolerant minimum broadcast networks.
\newblock {\em Networks}, 27, 1996.

\bibitem{BH94}
A.~Bagchi and S.~L. Hakimi.
\newblock Information dissemination in distributed systems with faulty units.
\newblock {\em IEEE Transactions on Computers}, 43(6):698--710, 1994.

\bibitem{BH86}
K.~A. Berman and M.~Hawrylycz.
\newblock Telephone problems with failures.
\newblock {\em SIAM Journal on Algebraic and Discrete Methods}, 7(1):13--17,
  1986.

\bibitem{BDP97}
P.~Berman, K.~Diks, and A.~Pelc.
\newblock Reliable broadcasting in logarithmic time with {Byzantine} link
  failures.
\newblock {\em Journal of Algorithms}, 22(2):199--211, 1997.

\bibitem{B73}
L.~A. Bjork.
\newblock Recovery scenario for a db/dc system.
\newblock In {\em ACM'73: Proceedings of the annual conference}, pages
  142--146, New York, NY, USA, 1973. ACM Press.

\bibitem{ChM84}
J.-M. Chang and N.~F. Maxemchuk.
\newblock Reliable broadcast protocols.
\newblock {\em ACM Transactions on Computer Systems}, 2(3):251--273, 1984.

\bibitem{CDP94}
B.~Chlebus, K.~Diks, and A.~Pelc.
\newblock Broadcasting in synchronous networks with dynamic faults.
\newblock {\em Networks}, 27, 1996.

\bibitem{CDP91}
B.~S. Chlebus, K.~Diks, and A.~Pelc.
\newblock Optimal broadcasting in faulty hypercubes.
\newblock In {\em FTCS}, pages 266--273, 1991.

\bibitem{C+88}
F.~R.~K. Chung, Z.~F\"uredi, R.~L. Graham, and P.~Seymour.
\newblock On induced subgraphs of the cube.
\newblock {\em Journal of Combinatorial Theory Series A}, 49:180--187, 1988.

\bibitem{D74}
E.~W. Dijkstra.
\newblock Self-stabilizing systems in spite of distributed control.
\newblock {\em Commun. ACM}, 17(11):643--644, 1974.

\bibitem{DP92}
K.~Diks and A.~Pelc.
\newblock Almost safe gossiping in bounded degree networks.
\newblock {\em SIAM Journal on Discrete Mathematics}, 5(3):338--344, 1992.

\bibitem{DKKS06}
S.~Dobrev, R.~Kr{\'a}lovi\v{c}, R.~Kr{\'a}lovi\v{c}, and N.~Santoro.
\newblock On fractional dynamic faults with threshold.
\newblock In P.~Flocchini and L.~Gasieniec, editors, {\em SIROCCO}, volume 4056
  of {\em Lecture Notes in Computer Science}, pages 197--211. Springer, 2006.

\bibitem{DV99}
S.~Dobrev and I.~Vr{\v t}o.
\newblock Optimal broadcasting in hypercubes with dynamic faults.
\newblock {\em Information Processing Letters}, 71(2):81--85, 1999.

\bibitem{DV04}
S.~Dobrev and I.~Vr{\v t}o.
\newblock Dynamic faults have small effect on broadcasting in hypercubes.
\newblock {\em Discrete Applied Mathematics}, 137(2):155--158, 2004.

\bibitem{D00}
S.~Dolev.
\newblock {\em Self-stabilization}.
\newblock MIT Press, Cambridge, MA, USA, 2000.

\bibitem{FLP85}
M.~J. Fischer, N.~A. Lynch, and M.~S. Paterson.
\newblock Impossibility of distributed consensus with one faulty process.
\newblock {\em Journal of the ACM}, 32(2):374--382, 1985.

\bibitem{F+97}
P.~Flocchini, B.~Mans, and N.~Santoro.
\newblock On the impact of sense of direction on message complexity.
\newblock {\em Information Processing Letters}, 63(1):23--31, 1997.

\bibitem{F+98}
P.~Flocchini, B.~Mans, and N.~Santoro.
\newblock Sense of direction in distributed computing.
\newblock In {\em International Symposium on Distributed Computing}, pages
  1--15, 1998.

\bibitem{F92}
P.~Fraigniaud.
\newblock Asymptotically optimal broadcasting and gossiping in faulty hypercube
  multicomputers.
\newblock {\em IEEE Transactions on Computers}, 41(11):1410--1419, 1992.

\bibitem{GP98}
L.~Gasieniec and A.~Pelc.
\newblock Broadcasting with linearly bounded transmission faults.
\newblock {\em Discrete Applied Mathematics}, 83(1--3):121--133, 1998.

\bibitem{HHL88}
S.~Hedetniemi, S.~Hedetniemi, and A.~Liestman.
\newblock A survey of broadcasting and gossiping in communication networks.
\newblock {\em Networks}, 18:319--349, 1988.

\bibitem{KKR03}
R.~Kr{\'a}lovi{\v c}, R.~Kr{\'a}lovi{\v c}, and P.~Ru{\v z}i{\v c}ka.
\newblock Broadcasting with many faulty links.
\newblock In J.~F. Sibeyn, editor, {\em SIROCCO}, volume~17 of {\em Proceedings
  in Informatics}, pages 211--222. Carleton Scientific, 2003.

\bibitem{LN00}
Z.~Liptak and A.~Nickelsen.
\newblock Broadcasting in complete networks with dynamic edge faults.
\newblock In F.~Butelle, editor, {\em OPODIS}, Studia Informatica Universalis,
  pages 123--142. Suger, Saint-Denis, rue Catulienne, France, 2000.

\bibitem{LPS80}
M.~Pease, R.~Shostak, and L.~Lamport.
\newblock Reaching agreement in the presence of faults.
\newblock {\em Journal of the ACM}, 27(2):228--234, 1980.

\bibitem{P91}
A.~Pelc.
\newblock Broadcasting in complete networks with faulty nodes using unreliable
  calls.
\newblock {\em Information Processing Letters}, 40(3):169--174, 1991.

\bibitem{PP05}
A.~Pelc and D.~Peleg.
\newblock Feasibility and complexity of broadcasting with random transmission
  failures.
\newblock In {\em PODC '05: Proceedings of the twenty-fourth annual ACM
  SIGACT-SIGOPS symposium on Principles of distributed computing}, pages
  334--341, New York, NY, USA, 2005. ACM Press.

\bibitem{SW90}
N.~Santoro and P.~Widmayer.
\newblock Distributed function evaluation in the presence of transmission
  faults.
\newblock In {\em SIGAL '90: Proceedings of the International Symposium on
  Algorithms}, pages 358--367, London, UK, 1990. Springer-Verlag.

\bibitem{SAA98}
I.~Stanoi, D.~Agrawal, and A.~E. Abbadi.
\newblock Using broadcast primitives in replicated databases.
\newblock In {\em ICDCS '98: Proceedings of the The 18th International
  Conference on Distributed Computing Systems}, pages 148--155, Washington, DC,
  USA, 1998. IEEE Computer Society.

\end{thebibliography}

\newpage
\begin{appendix}
\section{Appendix}
\setcounter{pako}{\thelemma}

This section contains the omitted technical parts.

\subsection{Complete Graphs}

\setnumber{lemma}{lm:grd}
\begin{lemma}
After two steps of the greedy algorithm, at least 
$$1+\min\left\{\frac{n}{2},(n-1)(1-\alpha)\right\}$$
vertices are informed.
\end{lemma}
\begin{proof}
In the first step the initiator sends $n-1$ messages. Let $l\ge2$ be the number of informed vertices after the first step.
In the second step, $l(n-1)$ messages are sent, and 
$\max\{n-2,\alpha l(n-1)\}$ of them are lost. We distinguish two cases:\\[2mm]

\noindent
{\bf Case 1:} $\alpha l(n-1)\le n-2$\\
In this case, at most $n-2$ messages are lost, i.e. at least $l(n-1)-n+2$ are delivered. 
Among those delivered, at most $l(l-1)$ could have been
sent to already informed vertices. Moreover, since each uninformed vertex has at most $l$ 
informed neighbors, we get that the
number of informed vertices is at least
$$l+\frac{l(n-1)-n+2-l(l-1)}{l}=n-\frac{n-2}{l}$$
Since $l\ge2$ we get that the number of informed vertices after the two steps is a least $\frac{n}{2}+1$.

\vskip 2mm
\noindent
{\bf Case 2:} $\alpha l(n-1)>n-2$\\
This time, at most $\alpha l(n-1)$ messages are lost. 
Using similar arguments, we get that the number of informed vertices is at least
$$l+\frac{l(n-1)(1-\alpha)-l(l-1)}{l}=1+(n-1)(1-\alpha)$$
\end{proof}

\subsection{Hypercubes}

\setnumber{lemma}{lm:Q:firststeps}
\begin{lemma}
After the first two steps of the algorithm, at least $\frac{1-\alpha}{2}(2d-1)$ vertices
have the information.
\end{lemma}

\begin{proof}
In the first step, the initiator sends $d$ messages. Since at most $d-1$ can be lost,
some $r>0$ of them are delivered. In the second step, the initiator sends again $d$ messages,
but at the same time, each of the informed vertices sends $d-1$ messages to all its neighbors
except initiator. Hence, $d+r(d-1)$ messages are sent in the second step. Let us distinguish
two cases:

If $d-1$ messages are lost, then $d+(r-1)(d-1)$ messages are delivered. $r$ messages from the
initiator can be delivered to the already informed vertices which leaves $d+(r-1)(d-1)-r$ messages
that enter uninformed vertices. Since at most $r$ messages can be destined to the same vertex,
The number of informed vertices after two steps is at least
$1+r+\frac{d+(r-1)(d-1)-r}{r}\ge(1/2)(2d-1)$

If $\alpha [d+r(d-1)]$ messages are lost, then $(1-\alpha)[d+r(d-1)]-r$ messages arrive into
uninformed vertices. Hence, there is at least $\frac{1-\alpha}{r}[d+r(d-1)]+r\ge\frac{1-\alpha}{2}(2d-1)$ informed vertices.
\end{proof}

\setnumber{lemma}{lm:HypercubePass}
\begin{lemma}
Consider a $d$-dimensional hypercube with $k$ non-informed vertices and
$h$ hyperactive arcs. Let  $\varepsilon\in(0,1)$ be an arbitrary constant, and let $k>X/(1-\varepsilon)$ or $h>X(d-1)$. 
Then in the second step of a simple
round at least $\beta(h+\partial(k))$ acknowledgements are delivered,
where $\beta=(1-\alpha)^2$.
\end{lemma}
\begin{proof}
Let $S$ be the set of informed vertices. In the first step of the round, $h+\partial(S)$ messages are sent.
Since the edge boundary of informed and uninformed vertices is the same,
 at least $h+\partial(k)$ messages are sent.
We prove that $\alpha(h+\partial(k))\ge d-1$, so in the first step at most $\alpha(h+\partial(k))$ messages are lost,
and at least $(1-\alpha)(h+\partial(k))$ of them are delivered. Next we prove that $\alpha(1-\alpha)(h+\partial(k))\ge d-1$,
so in the second step  at least $(1-\alpha)^2(h+\partial(k))$ messages are delivered. Since $1-\alpha<1$, it is sufficient
to prove that $\alpha(1-\alpha)(h+\partial(k))\ge d-1$.

If $h>X(d-1)$ then obviously $h+\partial(k)\ge X(d-1)$ and the statement holds. Next, let us consider the case when
$h>X/(1-\varepsilon)$. We distinguish three cases and prove that in each case $\partial(k)\ge X(d-1)$.

\vskip 2mm
\noindent
{\bf Case 1:} $k\le 2^{\varepsilon d}$\\
In this case it holds $\partial(k)\ge k(d-\lg k)\ge kd(1-\varepsilon)$. Since $k>X/(1-\varepsilon)$, we get
$\partial(k)\ge Xd$.

\vskip 2mm
\noindent
{\bf Case 2:} $2^{\varepsilon d}\le k\le 2^d\left(1-\frac{1}{\bm{e}}\right)$\\
In this case $\partial(k)\ge k(d-\lg k)\ge 2^{\varepsilon d}\left(d-d-\lg\left(1-\frac{1}{\bm{e}}\right)\right)
=2^{\varepsilon d}\lg\frac{\bm{e}}{\bm{e}-1}\ge 0.6\cdot2^{\varepsilon d}$. Since $X$ is constant, for large enough $d$ it holds
$\partial(k)\ge 0.6\cdot2^{\varepsilon d}>X(d-1)$.

\vskip 2mm
\noindent
{\bf Case 3:} $2^d\left(1-\frac{1}{\bm{e}}\right)\le k$\\
First, let us consider a function $f(x):=x(d-\lg x)$, for $x\in\left\langle 0,2^d\right\rangle$. Since
$f'(x)=d-1/\ln2-\lg x$, $f(x)$ is increasing for $x\in\left\langle
0,2^d/\bm{e}\right\rangle$ and decreasing for $x\in\left\langle
2^d/\bm{e}, 2^d\right\rangle$.

Obviously, the edge boundary of uninformed vertices $\partial(k)$ is the same as the edge boundary of
informed vertices $\partial(2^d-k)$. Hence, we get
$\partial(k)\ge f(2^d-k)$. Since $2^d-k\le 2^d\frac{1}{\bm{e}}$, the minimum of $f(2^d-k)$ is attained for
the minimal value of $2^d-k$. From Lemma~\ref{lm:Q:firststeps} we know that $2^d-k>\frac{1-\alpha}{2}(2d-1)$,
so $\partial(k)\ge f\left(\frac{1-\alpha}{2}(2d-1)\right)=\frac{1-\alpha}{2}(2d-1)\left(d-\lg\frac{1-\alpha}{2}(2d-1)\right)
=(1-\alpha)d^2-O(d\lg d)$. Hence, for large enough $d$ we get $\partial(k)\ge X(d-1)$.
\end{proof}

\setcounter{lemma}{\thepako}
\begin{lemma}
\label{lm:analysis}
Let $x\ge 2$. It holds that $\lg\frac{x+1}{x}\ge\frac{1}{x}$.
\end{lemma}
\begin{proof}
The statement is equivalent to:
$$\forall x\ge 2: \frac{1}{x}\ge 2^{\frac{1}{x}}-1$$
Substituting $y:=\frac{1}{x}$:
$$\forall y\in \left(0,1/2\right\rangle: y\ge 2^y-1$$
For $y=0$ the equality holds. Hence it is sufficient to prove that the
derivative of the left side is larger than the derivative of the right side
for $y\in (0,1/2\rangle$, i.e. $1\ge 2^y\ln 2$, which obviously holds.
\end{proof}

\setnumber{lemma}{lm:Q:part:I}
\begin{lemma}
After performing $O(d^2)$ simple rounds on a $d$-dimensional hypercube at
least $2^d/3$ vertices are informed.
\end{lemma}
\begin{proof}
Let $l:=2^d-k$ be the number of informed vertices and $b$ be the number of
passive arcs at the beginning of some simple round. Obviously $b\le ld$. Since the conditions of Lemma
\ref{lm:HypercubePass} are met, at least 
$\beta \partial(k)$ acknowledgements are delivered in one simple round.
Furthermore, the edge boundary of informed
vertices is also the boundary of uninformed vertices, so the number of delivered acknowledgements
is at least $\beta \partial(l)$. Because every delivered acknowledgement adds one passive arc,
the number of passive arcs grows at least to $b'=b+\beta \partial(l)$ after this round.

First, let us consider a function $f(x):=x(d-\lg x)$, for $x\in\left\langle 0,2^d\right\rangle$. Since
$f'(x)=d-1/\ln2-\lg x$, $f(x)$ is increasing for $x\in\left\langle
0,2^d/\bm{e}\right\rangle$ and decreasing for $x\in\left\langle
2^d/\bm{e}, 2^d\right\rangle$.

As $b/d\le l\le 2^d/3\le
2^d/e$ it holds that $\partial(b/d)\le\partial(l)$. Hence we have the following
lower bound on $b'$:
$$b'\ge b+\beta\partial\left(\frac{b}{d}\right)\ge
b+\beta\frac{b}{d}\left(d-\lg\frac{b}{d}\right)\ge
b\left(1+\beta\frac{d-\lg\frac{b}{d}}{d}\right)
$$
The lower bound on $b$ implies the inequality $\lg\frac{b}{d}\le
d+\lg(1/3))$. Hence it holds
$$b'\ge b\left(1+\beta\frac{-\lg(1/3)}{d}\right)=
b\left(1+\frac{1}{\frac{d}{\beta\lg 3}}\right)$$
We have shown that the number of passive arcs
grows exponentially with number of simple rounds performed. As it
can not grow above $d2^d/3$ without informing at least $2^d/3$ vertices,
we can estimate an upper bound on number of required simple rounds:
$$T\le\frac{\lg (d2^d/3)}{\lg\left(1+\frac{1}{\frac{d}{\beta\lg 3}}\right)}$$
For large enough $d$, Lemma \ref{lm:analysis} is applicable, hence
proving the Lemma:
$$T\le \lg (d2^d/3){\frac{d}{\beta\lg 3}}=O\left(d^2\right)$$

\end{proof}

\setnumber{lemma}{lm:Q:part:II}
\begin{lemma}
Let $\varepsilon\in(0,1)$ be an arbitrary constant, and let
$k_i\le(2/3)2^d$ be the number of uninformed vertices and $h_i$ be the
number of hyperactive arcs of a $d$-dimensional hypercube at the beginning
of round $i$. Then after $O(d^2)$ simple rounds there are at most 
$X/(1-\varepsilon)$ uninformed vertices and at most $X(d-1)$ hyperactive arcs.
\end{lemma}
\begin{proof}
Similarly to the proof of Theorem \ref{thm:AlmostBroadcast-Kn} let us consider the
measure $M_i:=2dk_i+h_i$.
Requirements of the Lemma ensure that $M_i\le O(d2^d)$. It is easy to
see that $M_i$ decreases with every acknowledgement delivered: if the
acknowledgement is delivered over a hyperactive arc, the value of $h_i$
decreases by 1. If it is delivered over an active arc, new vertex is
informed, hence the value of $k_i$ decreases by 1 and the value of $h_i$
increases by at most $2d-1$.

We show that the value of $M_i$ decreases by a certain multiplicative
factor in every simple round as long as the requirements of Lemma
\ref{lm:HypercubePass} hold. In one simple round at least
$\beta(h_i+\partial(k_i))$ acknowledgements are delivered, hence the value
$M_i$ decreases to at most:
\begin{eqnarray*}
M_{i+1}&\le& 2dk_i+h_i-\beta(h_i+\partial(k_i))\le h_i(1-\beta)+2dk_i-\beta k_i(d-\lg k_i)= \\
 & = & h_i(1-\beta) + 2dk_i\left(1-\beta+\beta\frac{\lg k_i}{d}\right) \end{eqnarray*}
Using the inequality $\lg k_i\le d+\lg(2/3)$ yields:
$$M_{i+1}\le h_i(1-\beta) + 2dk_i\left(1+\frac{\beta\lg (2/3)}{d}\right)$$
Hence for large enough $d$ it holds:
$$M_{i+1}\le (h_i+2dk_i)\left(1+\frac{\beta\lg (2/3)}{d}\right)$$

Since the requirements of the Lemma ensures that $M_i\le (7/3)d2^d$, the
requirements of Lemma \ref{lm:HypercubePass} can hold for at most
$$T:=\frac{\lg \left(\frac{7}{3}d2^d\right)}{\lg\left(1+\frac{1}{\frac{d}{\beta\lg
(2/3)}}\right)}$$
time steps. According to Lemma \ref{lm:analysis} for large $d$ it holds
that
$$T\le \lg \left(\frac{7}{3}d2^d\right) \frac{d}{\beta\lg
(2/3)} = O\left(d^2\right)$$
which concludes the proof.
\end{proof}

\subsection{Complete Broadcast}

\setnumber{theorem}{thm:K:sod}
\begin{theorem}
It is possible to perform broadcasting on complete graphs with chordal
sense of direction in time $O(\log n)$.
\end{theorem}
\begin{proof}
We present an algorithm for solving the broadcasting problem:

\begin{description}
\item[Phase 1] The algorithm from Lemma \ref{lm:AllButOne-Kn} is used.
This takes $O(\log n)$ time, all vertices but one are informed and the
vertex $0$ or the vertex $1$ knows a set $U$ of constant size containing
candidates for the uninformed vertex.

The rest of the algorithm is multiplexed into two parts, treating these two
cases separately. In the remaining of the description we assume that the vertex
$0$ knows the set $U$.

\item[Phase 2] The algorithm from Lemma \ref{lm:AllButOne-Kn} is used to
broadcast the set $U$, together with the original information, to all
vertices but one. This takes $O(\log n)$ time again. 

After the Phase 2 is finished, two cases are possible: Either the
uninformed vertex of the Phase 2 is different from or is the same as the
uninformed vertex of the Phase 1. In the former case all vertices are
informed. The rest of the algorithm handles the latter case.

\item[Phase 3] If not all vertices are informed, then there is a single
uninformed vertex $v$. Furthermore, every informed vertex knows the set $U$
of constant size such that $v\in U$. Every informed vertex iterates through
the set of $U$; in $i$-th time step of the current phase it sends the
message to $i$-th member of $U$. Eventually, the uninformed vertex is
processed. Since all $n-1$ informed vertices are
doing the same, exactly $n-1$ messages are sent to the uninformed vertex,
hence finishing the broadcast.
\end{description}

The time complexity of the Phase 1 and Phase 2 is $O(\log n)$; the time
complexity of the Phase 3 is $O(|U|)=O(1)$. Hence the algorithm correctly
solves the broadcasting on complete graphs in time $O(\log n)$. 
\end{proof}

\end{appendix}
\end{document}